\documentclass[fleqn,twoside]{article}
\usepackage{espcrc2}

\usepackage{amsmath, amsfonts, amsthm, amssymb, graphicx}
\usepackage{xspace}
\usepackage{epsfig}
\usepackage{psfrag}
\usepackage[usenames]{color}
\usepackage{hyperref}
\usepackage{cite}

\newcommand{\op}{\ensuremath{\mathcal{O}}\xspace}

\newcommand{\fdel}[2][]{\ensuremath{\frac{\delta #1}{\delta #2}}}

\newcommand{\vev}[1]{\ensuremath{\langle #1 \rangle}\xspace}

\def\ie{{\it i.e.\ }}

\newcommand{\be}{\begin{equation}}
\newcommand{\ee}{\end{equation}}
\def\ba{\begin{array}}
\def\ea{\end{array}}
\newcommand{\bea}{\begin{eqnarray}}
\newcommand{\eea}{\end{eqnarray}}
\newcommand{\ls}[1]{#1}

\begin{document}
\title{Real-time gauge/gravity duality and ingoing boundary conditions}
\author{Balt C. van Rees\address{ITFA, Valckenierstraat 65, 1018 XE Amsterdam, The Netherlands}}
\date{\today}
\begin{abstract}
In \cite{us,us2} a general prescription was presented for the computation of real-time correlation functions using the gauge/gravity duality. I apply this prescription to the specific case of retarded thermal correlation functions and derive the usual ingoing boundary conditions at the horizon for bulk fields. The derivation allows me to clarify various issues, in particular the generalization to higher-point functions and the relevance of including the regions beyond the horizon.
\end{abstract}

\maketitle

\subsection*{Introduction}
The gauge/gravity duality \ls{\cite{Maldacena:1997re,Gubser:1998bc,Witten:1998qj}} is by now firmly rooted in a well-developed dictionary between the gauge theory and the gravity side. Within the supergravity approximation, most entries in the dictionary can be conveniently summarized in the familiar formula
\be
\label{eq:zs} 
Z_{\text{qft}}[J] = \exp \big(-S_{\text{sugra}}[J]\, \big)\,,
\ee
with $J$ representing simultaneously QFT sources as well as boundary conditions for the supergravity fields.

However, \eqref{eq:zs} is really valid only in imaginary time. The real-time dictionary is necessarily more involved than the continuation of \eqref{eq:zs}, as it becomes necessary to specify initial and final field theory states on the left-hand side and initial and final supergravity data on the right-hand side. On general grounds one expects these boundary and bulk initial data to be related, and indeed a precise map was recently exposed in \cite{us,us2}. As we show in detail in \cite{us2}, this map directly yields a prescription for the computation of any real-time correlation function from a holographic background, just as concrete and generally valid as \eqref{eq:zs} in Euclidean signature.

Although the problem of initial data in gauge/gravity duality was never fully addressed before, specific prescriptions did exist in special cases. In particular, the authors of \cite{Son:2002sd} used a black hole argument to specify the initial and final bulk data in the case of retarded real-time thermal correlation functions: their prescription is to use `purely ingoing' boundary conditions for the supergravity fields at a bulk horizon. Subsequently, in \cite{Herzog:2002pc} these ingoing boundary conditions were tied to `natural' boundary conditions in the same way as retarded and time-ordered correlation functions are related in field theory. The recipe of \cite{Son:2002sd,Herzog:2002pc} turned out to yield consistent results and is by now widely used.

However, adhering only to this specific recipe would leave several questions unanswered. For example, in general one expects the field theory state (or rather ensemble) to determine all the initial conditions, including any boundary conditions for bulk fluctuations. How does this work precisely for this recipe? More specifically, can we change the state somewhat and obtain different (`non-natural') boundary conditions as well? And if the prescription is related to an on-shell action like \eqref{eq:zs} as suggested in \cite{Herzog:2002pc}, why can we ignore surface contributions to this action from the initial and final boundaries?

In this note I apply the general real-time prescription of \cite{us,us2} to the holographic computation of retarded thermal correlation functions and show that it reduces almost precisely to the recipe of \cite{Son:2002sd}. In the process, all the questions raised in the previous paragraph can be answered as well. In the last sections, I discuss the generalization to higher-point functions and some general lessons about the prescription.

\subsection*{A complete dictionary}
\begin{figure}
\psfrag{Im t}{Im $t$}
\psfrag{Re t}{Re $t$}
\psfrag{1}{$1$}
\psfrag{2}{$2$}
\psfrag{J1}{{\color{Maroon} $J_1$}}
\psfrag{J2}{{\color{Maroon} $J_2$}}
\centering
\includegraphics[width=5.5cm]{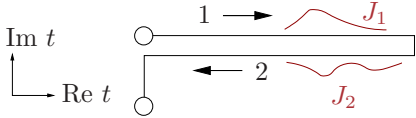}
\vskip -0.7cm
\caption{\label{fig:thermalcontour}A real-time thermal contour in the complex time plane. The circles should be identified. The two Lorentzian segments are labelled 1 and 2 on which we have sources $J_1$ and $J_2$, respectively.}
\vskip -0.7cm
\end{figure}
Consider a field theory at finite temperature $T = 1/\beta$. The dynamics of the corresponding gas or plasma is described by \emph{real-time thermal correlation functions}. These correlators can be obtained \cite{Landsman:1986uw} from a path integral along a contour in the complex time plane as sketched in Fig.~\ref{fig:thermalcontour}, with sources $J_1$ and $J_2$ (for an operator $\op$) on the two horizontal segments of the contour. We will in particular compute the \emph{retarded} correlator $i \Delta_R(x,x') =  \theta(t-t') \vev{[\op(x),\op(x')]}$, which is obtained by setting $J_1 = J_2 \equiv J$ and expanding the one-point function of $\op$ to first order in $J$:
\be
\label{eq:linearresponse} 
\delta_J \vev{\op(x)} = \int \! d^d x' \Delta_R(x,x') J(x') + \ldots
\ee
We will use this equation for $\Delta_R$ below.

\begin{figure}[b]
\psfrag{1}{$1$}
\psfrag{2}{$2$}
\psfrag{M1}{$M_1$}
\centering
\includegraphics[width=6.5cm]{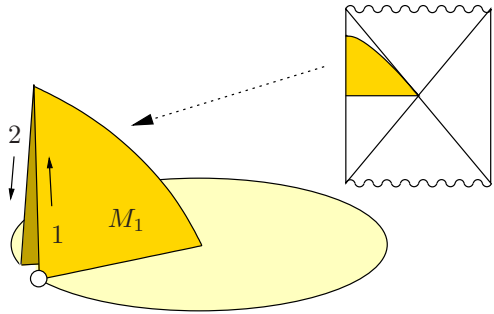}
\vskip -0.5cm
\caption{\label{fig:cutandpaste}The Euclidean segment of the contour is filled in with a disk; the two Lorentzian segments with two copies of a part of an eternal black hole.}
\end{figure}

Now let us apply the real-time gauge/gravity prescription of \cite{us,us2} for this specific field theory contour. The prescription instructs us to fill in the \emph{entire} field theory contour with bulk spacetimes. Consider therefore first the vertical segment in Fig.~\ref{fig:thermalcontour} and suppose that it can be filled in with a Euclidean black hole solution (see \cite{us2} for the case of thermal AdS). Topologically, this fills the imaginary time circle with a disk (plus some transverse space which is unimportant here). To add in the Lorentzian segments, we slice open the Euclidean black hole solution by making a cut in the disk, say at Euclidean time $\tau = 0$ up to the center of the disk. To the two cut surfaces we glue two copies of a segment of an eternal Lorentzian black hole solution which we will call $M_1$ and $M_2$. We finally glue $M_1$ and $M_2$ together along some late-time surface. The total space is sketched in Fig.~\ref{fig:cutandpaste}. I will discuss how this space is directly related to the eternal black holes of \cite{Maldacena:2001kr} in the last section below.

As usual, the sources $J_1,J_2$ on the boundary contour correspond to boundary data for the supergravity fields and switching them on causes perturbations on the background of Fig.~\ref{fig:cutandpaste}. According to \cite{us2}, these perturbations propagate from one segment to the other via certain \emph{matching conditions} that essentially guarantee $C^1$ continuity of the fields across the gluing. (The precise conditions can be derived from a saddle-point approximation.) Notice that the matching conditions can often be met by analytic continuation of the bulk solution, in which case they provide for the correct $i\epsilon$ insertions. The procedure for black holes is however more involved and I refer to \cite{us2} for details about the computations that follow.

\subsection*{Ingoing boundary conditions}
As we are interested in computing retarded correlation functions from  \eqref{eq:linearresponse}, we will have to consider a bulk perturbation on the background of Fig.~\ref{fig:cutandpaste} with $J_1 = J_2$. I will now show that precisely in these cases, the prescription of \cite{us,us2} yields ingoing boundary conditions for the bulk fields. As an example, I will consider a free bulk scalar field $\Phi$ with mass $m$ satisfying the bulk Klein-Gordon equation. For brevity, only the solution on $M_1$ will be written down below.

We assume that we can use separation of variables in $t$, the angular (or other transverse) coordinates $\vec \varphi$ and the radial coordinate $r$. One then finds four mode solutions,
\[
e^{-i\omega t} Y_l(\vec \varphi) \phi_{\pm \pm}(\omega,l,m^2,r)\,,
\]
with $Y_l$ some basis of harmonic functions on the transverse space. These modes are either purely ingoing ($\phi_{-+}$ and $\phi_{--}$) or purely outgoing ($\phi_{++}$ and $\phi_{+-}$); the second $\pm$ indicates the different possible analytic continuations across the horizons which we do not need here. Any solution $\Phi$ can be expanded in these modes with certain coefficients $a_{\pm\pm}(\omega,l)$:
\begin{multline}
\label{eq:Phi} 
\!\!\!\!\!\!\Phi(t,\vec \varphi,r) = \sum_{l} \int\!\! d\omega\, e^{-i\omega t} Y_l(\vec \varphi) (a_{++}\phi_{++} \\ + a_{+-}\phi_{--} + a_{-+}\phi_{-+} + a_{--} \phi_{--})\,.
\end{multline}

Now consider the solution corresponding to a delta-function source at $(t,\vec \varphi) = 0$ on $\partial_r M_1$ and denote the corresponding solution on $M_1$ as $\Delta_{[11]}$. If the modes are appropriately normalized, then $\Delta_{[11]}$ has the form \eqref{eq:Phi} with
\[
\begin{split}
 &a_{[11]++} = (1 - e^{\beta \omega})^{-1} \qquad  a_{[11]+-} = 0\\
 &a_{[11]-+} = 0 \qquad  a_{[11]--} = (1 - e^{- \beta \omega})^{-1}\,.
\end{split}
\]
These $a_{[11]\pm \pm}$ are uniquely determined by demanding normalizability along the radial boundary of the entire manifold of Fig.~\ref{fig:cutandpaste} (except of course at the origin of $\partial_r M_1$), combined with the matching conditions between the segments. This solution precisely satisfies the `natural' boundary conditions of \cite{Herzog:2002pc}.

Let us now move the delta-function source to the origin of $\partial_r M_2$. The perturbation propagates to $M_1$ via the matching conditions, where we denote the corresponding solution as $\Delta_{[21]}$. Its mode expansion has the coefficients:
\[
\begin{split}
&a_{[21]++} = (e^{\beta \omega}-1)^{-1} \qquad  a_{[21]+-} = 0\\
&a_{[21]-+} = (1 - e^{\beta \omega})^{-1} \qquad  a_{[21]--} = 0\,,
\end{split}
\]
which are obtained in the same way as the $a_{[11]}$.

With obvious modifications, $\Delta_{[11]}$ and $\Delta_{[21]}$ can be made into bulk-boundary propagators. We can then integrate them against a source $J_{1}$ on $\partial_r M_1$ and another source $J_{2}$ on $\partial_r M_2$. Adding the two solutions gives the \emph{unique} solution for given $J_1$ and $J_2$ that satisfies the matching conditions. In particular, if $J_1 = J_2$ then the bulk solution $\Phi_{1}$ on $M_1$ becomes:
\be
\label{eq:ingoingphi} 
\begin{split}
\Phi_{1} &= \int_{\partial_r M_1} \!\!J_1 \, \Delta_{[11]} + \int_{\partial_r M_2} \!\!J_2 \, \Delta_{[12]}\\
&= \int_{\partial_r M} \!\!J \, (\Delta_{[11]} + \Delta_{[12]})\,.
\end{split}
\ee
The last line is important, as it shows that the coefficients of $\Phi_{1}$ are alternatively given by a \emph{new} bulk-boundary propagator which is found by adding the two propagators given above. It has the coefficients
\[
a_{[1]\pm \pm} = a_{[11]\pm \pm} +  a_{[12]\pm \pm}
\]
and we find that all the outgoing modes $a_{[1]+\pm}$ vanish. Therefore, the free-field bulk solution for $J_1 = J_2$ satisfies purely ingoing boundary conditions.

This solves two of the above puzzles: the choice between natural or ingoing conditions \emph{is} determined by field theory, namely by switching on the source $J_2$ as well. We can also change the ensemble by adding sources on the Euclidean segment; the gluing and matching ensure that this indeed leads to `non-natural' conditions at the horizon.

\subsection*{On-shell action and correlation functions}
Let me now compute a retarded correlator. The prescription of \cite{us,us2} is to use \eqref{eq:zs} with a factor $i$, but on the right-hand side we have to sum the on-shell supergravity actions for all the segments in Fig.~\ref{fig:cutandpaste}. The action for each segment contains initial and final surface terms that arise from the integration by parts in both the radial and the time coordinates. However, it is precisely by virtue of the matching conditions that all these terms cancel between adjacent segments, which nicely resolves the final question of the introduction as well.

Using a Fefferman-Graham radial coordinate $r$, we find that $\Phi_1$ has an expansion near $\partial_r M_1$ of the form:
\be
\!\!\!\!\!\!\!\!\!\Phi_1 = J(x) r^{\Delta - d} + \ldots + \phi_{(2\Delta - d)}(x) r^{-\Delta} + \ldots
\label{eq:fgexpansion} 
\ee
with $x = (t,\vec \varphi)$ and $m^2 = \Delta(\Delta -d)$. After renormalization \cite{Skenderis:2002wp} and differentiation, the dual one-point function of an operator on $\partial_r M_1$ in the presence of sources is given by the normalizable term,
\be
\label{eq:vevholographic} 
\vev{\op(x)} = - (2\Delta - d) \phi_{(2\Delta-d)}(x)\,,
\ee
plus possible contact terms. We can then use \eqref{eq:linearresponse} to obtain the retarded two-point function
\[
\Delta_R(x,x') = - (2\Delta -d) \fdel[\phi_{(2\Delta-d)}(x)]{J(x')}\,,
\]
with $\phi_{(2\Delta-d)}$ the normalizable term in the purely ingoing solution $\Phi_{1}$. Up to the normalization which arises from \cite{Skenderis:2002wp}, this is precisely the prescription of \cite{Son:2002sd,Herzog:2002pc}.

\ls{ 
\subsection*{Higher-point correlation functions}
The previous discussion applied to two-point functions only. In this section, I show that ingoing boundary conditions can be used to compute specific three- and higher-point correlation functions as well.

Higher-point correlators cannot be computed holographically from a free-field analysis and the full, nonlinear bulk field equations have to be considered instead. These can be solved perturbatively, for example by using a bulk-bulk propagator. For a background as in Fig.~\ref{fig:cutandpaste}, which consists of multiple segments, such a bulk-bulk propagator has to be defined for the entire manifold \cite{us2}. Its form is again uniquely determined by demanding normalizability along the entire radial boundary, combined with the matching conditions between the segments.

As an example, consider the corrections arising from a nonlinear term $\sim \lambda \Phi^2$ in the Klein-Gordon equation. I will again focus on the bulk solution $\Phi_1$ on $M_1$ and consider only the case $J_1 = J_2$.

The first-order correction to $\Phi_1$ is obtained by integrating the bulk-bulk propagator against the square of the free-field solution on all the segments. For $J_1 = J_2$, the free-field solution is causal (like the boundary response) and in particular vanishes on the Euclidean segment, so we may restrict the integration to $M_1$ and $M_2$ only. An argument along the same lines as above then shows that this integral over $M_1$ and $M_2$ can be rewritten as a \emph{single} integral over $M_1$ with a \emph{new} bulk-bulk propagator, namely precisely one that satisfies purely ingoing boundary conditions. 

Therefore, as far as $\Phi_1$ is concerned, we may forget about the backward-going segment $M_2$ altogether and use purely ingoing boundary conditions for the bulk-bulk propagator instead. This result extends to all orders in $\lambda$ and also holds if one uses for example a derivative expansion, as long as the bulk perturbation is causal.

On the other hand, $M_2$ \emph{is} important for the boundary theory, since we need the full boundary contour to understand precisely which correlators we are computing. In particular, the solution near $\partial_r M_2$ shows that purely ingoing bulk solutions actually correspond to $J_1 = J_2$ on the boundary.

As an example, consider a holographically computed three-point function:
\be
\label{eq:delta} 
\!\!\!\!\!\!\!\!\Delta(x,x',x'')  = (2\Delta -d) \frac{\delta^2 \phi_{(2\Delta-d)}(x)}{\delta J(x'') \delta J(x')}\,,
\ee
with $\phi_{(2\Delta-d)}(x)$ again the normalizable component of $\Phi_1$, which is now a purely ingoing (approximate) solution to the nonlinear field equations. To find the precise field theory expression for $\Delta(x,x',x'')$, we need to expand the right-hand side of \eqref{eq:linearresponse} to quadratic order in $J = J_1 = J_2$. Following \cite{Landsman:1986uw}, the quadratic term is:
\begin{align}
\!\!\!\!\!\!\!\!\!\!\!\!\!\delta_J \vev{\op(x)} &= \ldots - \int \! d^d x' \int\! d^d x'' \Delta_{RR}(x,x',x'') \nonumber \\
& \qquad \times J(x') J(x'') + \ldots \label{eq:deltarr}
\end{align}
with $\Delta_{RR}$ the retarded three-point function,
\[
\!\!\!\!\!\!\!\!\!\!\Delta_{RR} = \theta(t-t')\theta(t-t'') \big \langle \big[ [\op(x),\op(x')],\op(x'') \big] \big\rangle\,.
\]
Therefore, the three-point function \eqref{eq:delta} computed using purely ingoing boundary conditions is:
\[
\!\!\!\!\!\!\!\!\!\!\!\Delta(x,x',x'') = \Delta_{RR}(x,x',x'') + \Delta_{RR}(x,x'',x')\,.
\]
This argument readily generalizes to higher-point functions.
}
\subsection*{Conclusion}
I have shown that the prescription of \cite{us,us2} reduces to the ingoing prescription of \cite{Son:2002sd,Herzog:2002pc} when $J_1 = J_2$, \ie when one computes retarded correlation functions. \ls{I also discussed which higher-point Lorentzian correlation functions are computed with ingoing boundary conditions. }Along the way, the questions raised in the introduction were answered as well.

This example illustrates that a bulk solution for the \emph{entire} field theory contour is both necessary and sufficient to make the real-time dictionary just as precise as its imaginary-time counterpart.

\subsection*{Discussion}
In this final section, I will discuss the construction of Fig.~\ref{fig:cutandpaste} in more detail.

This manifold is a variation of the one presented in \cite{Maldacena:2001kr}, where the initial state for an eternal Lorentzian black hole was given by half a Euclidean black hole. As shown in \cite{us2}, the prescription relating states and Euclidean manifolds can be made precise if one includes a \emph{second} copy of the manifold of \cite{Maldacena:2001kr}, which should be glued to the first copy along some late-time hypersurface. In field theory, this second copy corresponds to the backward-going segments of the contour and the specification of the \emph{final} rather than the initial state. The two copies are in this case identical, precisely because the initial and final field theory states are identical as well.

A distinguishing feature of this construction is that one may freely deform the late-time surface between the two Lorentzian segments without affecting boundary correlators \cite{us2}. In particular, the manifold of Fig.~\ref{fig:cutandpaste} can be directly obtained from the space of the previous paragraph by moving the late-time surface so far down that some Lorentzian parts of the space completely disappear.

This deformation freedom for the late-time hypersurface is reminiscent of field theory, where it is not necessary to path integrate further than the latest operator insertion either. It shows that one does not need to compute the complete real-time development of a spacetime to compute real-time correlators. For instance, there is no need to include the singularity or new asymptotic regions beyond a possible inner horizon. No significance is attached to any horizon either, as the final surface may be deformed freely through such horizons. In particular, any construction invoking membranes at the horizon is at most an effective (but interesting) description.

One may think that the deformation freedom contradicts the fact that the entire Lorentzian spacetime is encoded in boundary correlators. This contradiction disappears if one realizes that the boundary correlators in principle have the power to resolve the field theory states. In the bulk, these states completely determine the Euclidean segments which in turn provide sufficient initial data to reconstruct the Lorentzian spacetime as well. The spacetime is therefore encoded completely in the real-time correlators, although sometimes only in an indirect way.

\ls{\subsection*{Acknowledgments}
I would like to thank Kostas Skenderis for collaboration on \cite{us,us2}, Sheer El-Showk and Dam Thanh Son for discussions, and the organizers of the 2008 Carg\`ese Summer School for an inspiring school.}

\providecommand{\href}[2]{#2}\begingroup\raggedright\endgroup

\end{document}